\documentstyle[epsfig,12pt]{article}
\begin{document} 
%
%
%
%
\newcommand{\x}{\cdot}
\newcommand{\ra}{\rightarrow}
\newcommand{\pom}{\mbox{$\rm{\cal P}$omeron}}
\newcommand{\flux}{\mbox{$ F_{{\cal P}/p}(t, \xi)$}}
\newcommand{\ap}{\mbox{$\bar{p}$}}
\newcommand{\pap}{\mbox{$p \bar{p}$}}
\newcommand{\SPS}{\mbox{S\pap S}}
\newcommand{\xp}{\mbox{$x_{p}$}}
\newcommand{\xf}{\mbox{$x_{F}$}}
\newcommand{\et}{\mbox{${E_T}$}}
\newcommand{\etj}{\mbox{$\et ^{jet}$}}
\newcommand{\sumet}{\mbox{$\Sigma \et$}}
\newcommand{\sumetj}{\mbox{$\Sigma \et ^{jet}$}}
\newcommand{\pt}{\mbox{${p_t}$}}
\newcommand{\mpr}{\mbox{${ m_p}$}}
\newcommand{\mpi}{\mbox{${ m_\pi}$}}
\newcommand{\rs}{\mbox{$ \sqrt{s}$}}
\newcommand{\rsp}{\mbox{$ \sqrt{s'}$}}
\newcommand{\rsps}{\mbox{$ \sqrt{s} = 630 $ GeV}}
\newcommand{\lum}{\mbox{$\int {\cal L} {\rm dt}$}}
\newcommand{\T}{\mbox{$t$}}
\newcommand{\abt}{\mbox{${ |t|}$}}
\newcommand{\di}{\mbox{d}}
\newcommand{\sigdifjets}{\mbox{$ \sigma_{sd}^{jets}$}}
\newcommand{\sigpomjets}{\mbox{$ \sigma_{{\cal P}p}^{jets}$}}
\newcommand{\sigdiftot}{\mbox{$ \sigma_{sd}^{total}$}}
\newcommand{\sigpomtot}{\mbox{$ \sigma_{{\cal P}p}^{total}$}}
\newcommand{\dsig}   {\mbox{$ {{ d^2 \sigma        }\over{d \xi dt}} $}}
\newcommand{\dsigdif}{\mbox{$ {{ d^2 \sigma _{sd}  }\over{d \xi dt}} $}}

\newcommand{\alam}{\mbox{$\overline{\Lambda}^{\circ}$}}
\newcommand{\lam}{\mbox{$\Lambda^{\circ}$}} 
\newcommand{\PRET}{\mbox{\Proton-\sumet}}
\begin{titlepage}
\begin{center}
{\large   }
{\large EUROPEAN ORGANIZATION FOR NUCLEAR RESEARCH}
\end{center}
\vspace{2 ex}
\begin{flushright}  
{
11 December, 1997
}
\end{flushright}
 
  
\begin{center}
{
\LARGE\bf
\rule{0mm}{7mm} Measurements of Inclusive {\boldmath \alam} Production \\
\rule{0mm}{7mm} with Large {\boldmath \xf} at the {\boldmath \SPS}-Collider
}\\
\vspace{4ex}

A. Brandt$^{1}$, S. Erhan, A. Kuzucu$^{2}$, M. Medinnis$^{3}$,\\
N. Ozdes$^{2}$, P.E. Schlein, M.T. Zeyrek$^{4b}$, J.G. Zweizig$^{5}$\\
University of California$^{*}$, Los Angeles, California 90024, USA. \\
\vspace{2 ex}
J.B. Cheze$^{a}$, J. Zsembery \\
Centre d'Etudes Nucleaires-Saclay, 91191 Gif-sur-Yvette, France.
\end{center}
\vspace{2 ex}

\centerline{(UA8 Collaboration)}

\vspace{1 ex}

\begin{abstract}

We report results of inclusive measurements of \alam s, produced
in the forward direction at the \SPS\ with \rsps , using the 
UA8 small angle Roman Pot spectrometers.
These measurements cover the range in Feynman-\xf\ and transverse
momentum, $0.6 < \xf < 1.0$ and $0.4 < \pt < 0.7$ GeV, respectively. 
Within a systematic uncertainty of $\pm 20\%$ on the absolute
cross section measurements, the results are indistinguishable from those made 
by some of us in two earlier experiments at the CERN Intersecting Storage Rings,
with energies, $\sqrt{s} = 53 \, \rm and \,  62$~GeV.
In the \xf -range, 0.6-0.9, the absolute cross sections are lower 
by a factor of 2 to 3 
than the predictions of the Lund model as embodied
in the PYTHIA 5.6 and JETSET 7.3 Monte Carlo programs, indicating
inadequacies in knowledge of the baryon fragmentation function. 
For the largest
\xf -range, 0.9-1.0, the measurements agree with the Monte Carlo predictions.
We have measured the average \alam\ polarization for our events and
find ($6 \pm 12\%$), consistent with previous measurements at the ISR in 
the present region of \xf -\pt .

\end{abstract}

\begin{center}
Submitted to Nuclear Physics B \\
\end{center}
\vspace{1 ex}
\rule[.5ex]{16cm}{.02cm}
$^{*}$ Supported by U.S. National Science Foundation
Grant PHY94-23142 \\
$^{a}$ email:  cheze@mail.cern.ch \\
$^{b}$ email:  zeyrek@rorqual.cc.metu.edu.tr \\
$^{1}$ Now at Fermi National Accelerator Laboratory, Batavia, Illinois, 
U.S.A. \\ 
$^{2}$ Visitor from Cukurova University, Adana, Turkey; also supported by 
ICSC - World Lab.\\
$^{3}$ Present address: DESY, Zeuthen, Germany \\
$^{4}$ Visitor from Middle East Tech. Univ., Ankara, Turkey; supported by 
TUBITAK. \\
$^{5}$ Present address: DESY, Hamburg, Germany  \\
\end{titlepage}      

\pagebreak
\setlength{\oddsidemargin}{0 cm}
\setlength{\evensidemargin}{0 cm}
\setlength{\topmargin}{0.5 cm}
\setlength{\textheight}{22 cm}
\setlength{\textwidth}{16 cm}
\setcounter{totalnumber}{20}
\clearpage\mbox{}\clearpage
\pagestyle{plain}
\setcounter{page}{1}

\tableofcontents


\section{Introduction}
\label{intro}
\indent

Forward \lam\ and \alam\ production has been extensively studied at the
CERN Intersecting Storage Rings in $pp$ and $p\bar{p}$ 
interactions with large aperture spectrometers[1-8].
Most of the previous work has used inclusively measured \lam 's in the
reactions:
\begin{equation}
p \, \, + \, \, p \, \, \ra \, \, \lam \, \, + \, \, X
\label{eq:laminclusive}
\end{equation}
\begin{equation}
\ap \, \, + \, \, p \, \, \ra \, \, \alam \, \, + \, \, X
\label{eq:alaminclusive}
\end{equation}
although results on \lam 's produced diffractively\cite{pomquark,henkes} in 
$pp$ interactions have also been reported. 

\lam 's measured inclusively in the fragmentation region (large \xf )
of React.~\ref{eq:laminclusive} are found to satisfy Feynman scaling
throughout the ISR energy range. In addition, 
detailed measurements of the \lam\ 
polarization in the fragmentation region of
React.~\ref{eq:laminclusive} yielded values as large
as 40\%, and as large as 60\% for diffractively-produced \lam 's, where
there is no dilution from $\Sigma$ and $\Sigma ^{*}$ decay. 
A remarkable demonstration of the $s$-independence of the 
polarization\cite{sherwood} 
in the fragmentation region of 
React.~\ref{eq:laminclusive} is that it changes by $(0.8 \pm 1.7)\%$ 
between 31 and 62 GeV center-of-mass energy.

Tests of Feynman scaling were extended, for charged particles, from
the ISR to an \SPS -Collider
energy of 546 GeV by the UA5 Collaboration\cite{ua5}.
They find that, for charge particle production in the beam fragmentation
region, Feynman scaling over this larger energy range 
is valid to $\sim \pm10\%$.

In the work reported here, we have extended Feynman scaling tests 
over this energy range to
\alam 's produced in the beam fragmentation region of 
React.~\ref{eq:alaminclusive} with 630 GeV center-of-mass energy.
While Feynman-scaling is found to be valid over this large energy 
range -- the invariant cross section is shown to be almost indistinguishable
between ISR and SPS-collider energies -- the cross sections are found to be 
a factor of 2-3 times smaller than predicted by the
PYTHIA Monte-Carlo event generator.
This observation will allow PYTHIA to be ``tuned" to give more
reliable estimates of baryon production in the beam fragmentation region.
A measurement of \alam\ polarization in React.~\ref{eq:alaminclusive} 
is consistent with zero, in agreement with earlier measurements in
React.~\ref{eq:laminclusive} in the same \xf -\pt\ region.
 
The $\ap \pi^+$ 
decay products of the \alam\ produced in React.~\ref{eq:alaminclusive}
were both detected in one of the
two UA8 Roman pot spectrometers\cite{ua8hard}, installed above and below
the beam, respectively, in the direction of the outgoing antiprotons. 
Although the 
UA8 apparatus was not designed for this purpose, 
adequate acceptance was found for \alam 's which decay between 19 and 21 
meters downstream of the interaction point.
When the 
\SPS\ luminosity reached values over $10^{30}$ cm$^{-2}$sec$^{-1}$, 
the measurement of forward \alam 's 
in the \ap -arm became feasible. 

The two Roman pot spectrometers provide independent
measurements of React.~\ref{eq:alaminclusive}.
Having two independent spectrometers 
(above and below the circulating beams)
was particularly important in the early \lam\ polarization 
measurements\cite{erhanpol,sherwood}, made with
large forward multi-particle spectrometers operated
in two experiments at the CERN Intersecting Storage Rings.

\section{Apparatus and Trigger}

A detailed description of our apparatus, its properties, triggering
capabilities and interface to the UA2 experiment are given 
elsewhere\cite{ua8hard}. Thus, we only provide here a brief summary
of the spectrometer, in particular as regards the measurement\cite{thesis} of 
React.~\ref{eq:alaminclusive}.
UA8 was installed in the same interaction region
as the UA2 experiment\cite{ua2hard} and was interfaced to the UA2 data
acquisition system in such a way that UA8 could record data in parallel
with UA2.

The Roman-pot spectrometers, which used the low-$\beta$ machine quadrupole
magnets, consisted of four pot installations positioned 
in each arm of intersection LSS4 at the CERN \SPS -Collider.
The four spectrometers are labeled according to which arm they are in
(P for outgoing proton and M for outgoing antiproton),
and whether they are above or below the beam pipe (``U'' for ``UP'' and ``D'' 
for ``DOWN''). 
As discussed below, the MU and MD spectrometers were instrumented to allow
data from React.~\ref{eq:alaminclusive} to be recorded inclusively with the
spectrometers running in a stand-alone mode (i.e. trigger provided only
by the UA8 spectrometers and no UA2 detectors used in the analysis).

\begin{figure}
\begin{center}
\mbox{\epsfig{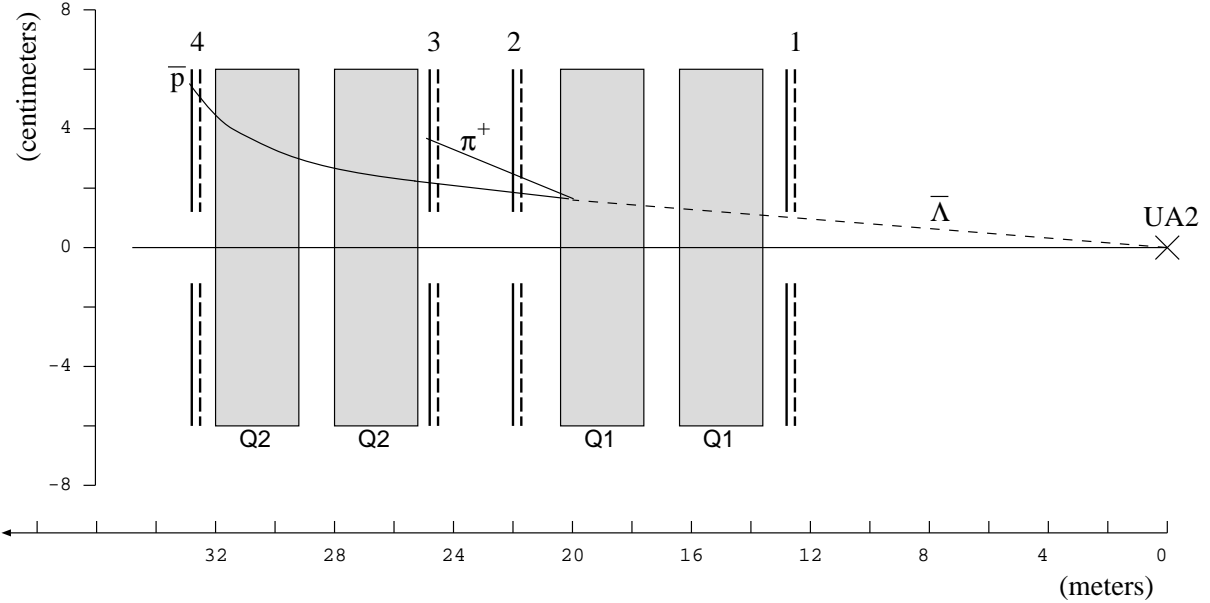}}
\end{center}
\caption[]{
General layout of the two UA8 spectrometers 
installed in the outgoing \ap -arm with a typical
\alam\ event, shown decaying in the vertical plane,
in the UP spectrometer (drawn with different vertical and horizontal scales).
The horizontal axis is the longitudinal $z$-axis. The vertical axis is the
$y$-coordinate.
Q1 and Q2 are each pairs of quadrupole magnets, The four chamber
packs (solid vertical lines) each consist of six wire planes as described 
in Ref.~\protect\cite{ua8hard}, are positioned before and after each
magnet at the longitudinal positions shown. 
Scintillation counters (dashed vertical lines) are installed next to each 
chamber pack. 
} 
\label{fig:lambda_layout}
\end{figure}

Figure~\ref{fig:lambda_layout} 
shows a side view of the two UA8 spectrometers used.
Each spectrometer includes
four Roman pots, each of which contain a chamber assembly 
and a plastic scintillator, Sc1-Sc4, used for trigger purposes. The
two quadrupoles Q1 and Q2 are, respectively, vertically
focusing and defocusing for a \ap\ coming from the
beam-crossing region. 

A system
of wire chambers was constructed with high bandwidth readout electronics.
As discussed in Ref.~\cite{ua8hard}, 
each chamber contained 6 multi-wire-proportional wire planes, 
$y, y', u, u', v$ and $v'$, with 28 wires each, spaced by 2 mm. 
$y$ and $y'$ have horizontal wires, $u, u'$ and $v, v'$ are stereo views tilted 
by $\pm7^{\circ}$, respectively, with respect to the horizontal. 
The primed planes are shifted by half
a wire spacing with respect to the unprimed planes, 
to solve the left-right ambiguity.
Thus, the $y$-resolution (vertical or bending plane) 
is much better than the $x$-resolution (horizontal).
With a 4-bit time-to-digital converter (5~ns least count)
on each wire, a chamber provides resolutions of
$\sim 65 \, \mu$m and $\sim 670 \, \mu$m
in the $y$ and $x$-views, respectively.
The readout electronics also provided 8-bit logic signals, with each bit
corresponding to an ``OR'' of four contiguous wires from a 32-channel
TDC\cite{ua8hard}.

\begin{figure}
\begin{center}
\mbox{\epsfig{file=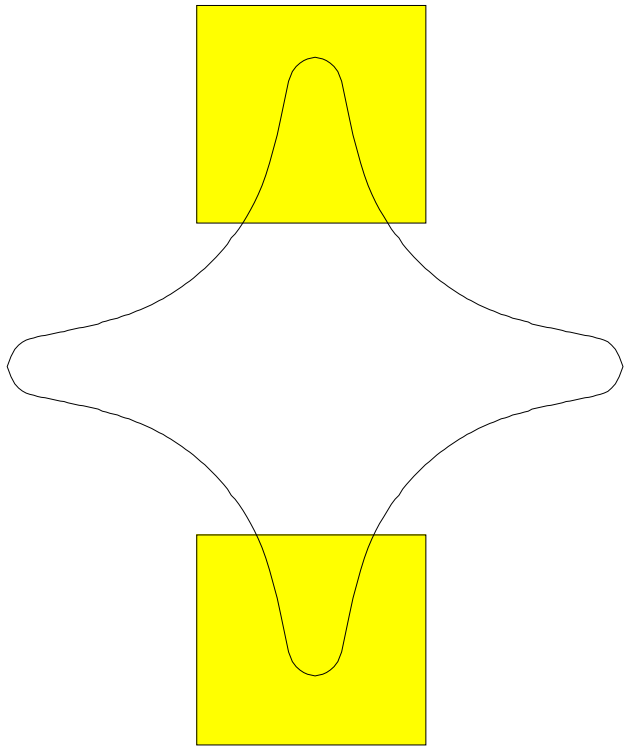,width=12.7cm}}
\end{center}
\caption[]{
The UA8 spectrometer aperture viewed from the interaction region. 
The shaded rectangles indicate the sensitive regions of the first wire chambers
at a distance $z$ = 13 m from the interaction region center.
The curved line indicates the walls of the beam pipe inside the quadrupole 
magnets.
}
\label{fig:aperture}
\end{figure}

Figure~\ref{fig:aperture} shows a ``beams-eye'' view of the UA8 chamber
aperture which is closest to the center of the interaction region.
The 4-lobed curve in the figure illustrates the contour of the 
beam pipe which follows that of the quadrupole-magnet pole pieces.
The overlap between the beam pipe and rectangular chambers above and below
the beam illustrate the limited azimuthal range through which a
particle may pass.
A discussion of the acceptance is given below in Sect.~\ref{geomacc}.
Data were recorded with the bottom edge of each pot set at
either 12 or 14 beam widths from the beam axis, referred to below
as the ``12$\sigma$ and 14$\sigma$'' data.

If a track is in the DOWN spectrometer, we define the ``adjacent" 
spectrometer to mean the UP spectrometer of the same arm. 
Similarly, we define ``opposite'' spectrometer to be the one 
diagonally opposite the one containing a trigger particle.

As shown in Fig.~\ref{fig:lambda_layout},
a typical \alam\ decays close to the end of the quadrupole Q1 in a 2 meter 
long region which starts 19 m from the interaction region. 
The \ap\ and $\pi^+$ are detected in chambers 2 and 3 
which are in a magnetic-field-free region between the two quadrupoles. The \ap\
travels through quadrupole Q2 and is detected in chamber 4 
at the end of the spectrometer. The low momentum $\pi^+$ is 
strongly deflected by quadrupole Q2 and does not reach the last 
chamber.
 
\vspace{2 ex}
\noindent {\large \bf \boldmath \alam\ Trigger}
\vspace{2 ex}

\begin{figure}
\begin{center}
\mbox{\epsfig{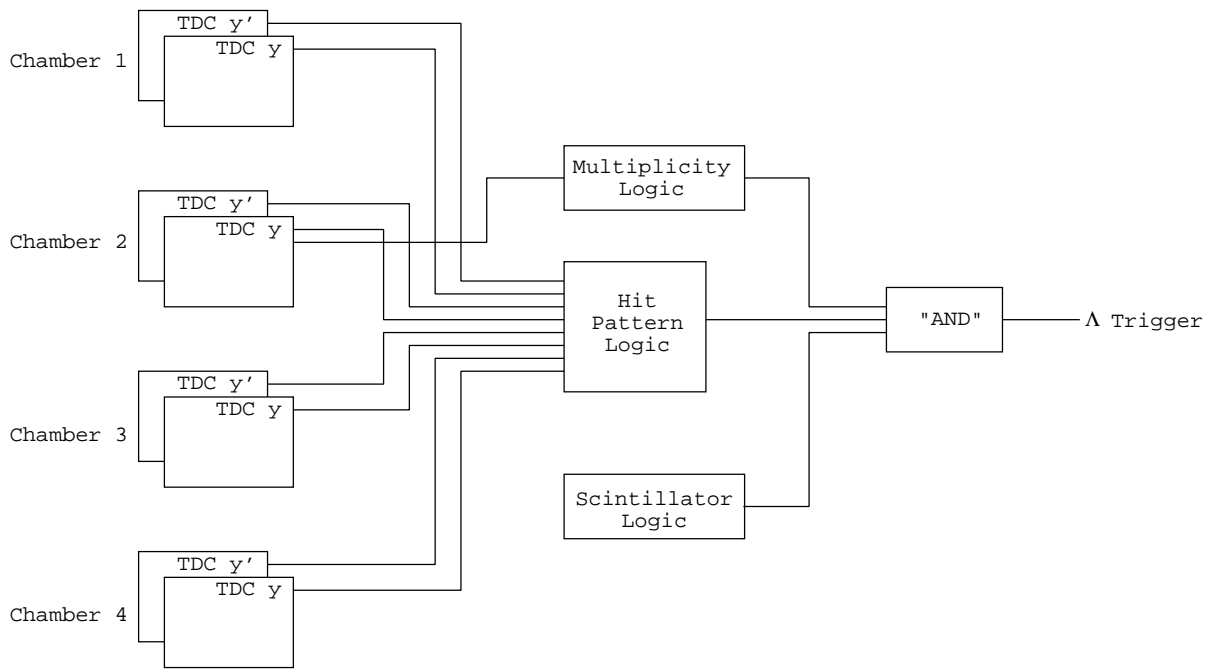}}
\end{center}
\caption[]{
Block diagram of the \alam\ trigger.
See text and Ref.~\protect\cite{ua8hard} for a detailed explanation.
}
\label{fig:lambda_tdc}
\end{figure}

The decay products of an \alam\ which decays
near the end of the first quadrupole, Q1, leave recognizable 
signatures in chambers 2, 3 and 4.  
Figure~\ref{fig:lambda_tdc} shows a block diagram for the \alam\ trigger for one
of the spectrometers. As shown in this figure, the \alam\ trigger had
three components: (1) hit pattern logic based on latched ``hit
groups"\footnote{A ``hit group" is a group of 4 contiguous wires, 
at least one of which has a hit signal on it.} in the $y$ 
and $y'$ planes of all
chambers in a spectrometer, (2) logic on the latched signals from the
four scintillators in a spectrometer and (3) a multiplicity signal
which was generated on the TDC card for the $y$ plane of chamber 2. A
more detailed description is given in the following sections.

\vspace{2 ex}
\noindent {\bf Hit pattern logic}
\vspace{2 ex}
 
For accepted events, the \ap\ traverses chambers 2, 3 and 4, while the $\pi ^+$
only traverses chambers 2 and 3. 
They are separated by at most 16 mm 
in chamber 2 and by at least 16 mm in chamber 3.
An efficient trigger was devised which required the following hit patterns.
For this purpose, a ``hit'' is defined as a latched anode signal from at
least one wire in a group of four successive wires
and a ``gap'' is defined when there are no latched anode
signals in any of a group of four successive wires. The groups are
numbered starting from the group closest to the beam.
 
\begin{itemize}
 
\item No hits in chamber 1,
 
\item chamber 2 not required to resolve the two tracks. Specifically,
chamber 2 must have
1 hit, 2 or 3 adjacent hits, or 2 hits separated by one gap in the
lowest three 4-wire groups and no hits elsewhere,
 
\item chamber 3 must resolve the two tracks. That is, we require
2 non-adjacent hits, or 2 adjacent hits and 1 non-adjacent hit,
 
\item chamber 4 can have only one track. Therefore we require,
1 hit in groups 3 through 7 and no hits in groups 1 and 2.
 
\end{itemize}
 
The logic for this scheme was implemented 
using a total of 5 CMOS PALs
for each spectrometer. The hit pattern decision took about 50 nsec.

\vspace{2 ex}
\noindent {\bf Scintillator logic}
\vspace{2 ex}
 
The scintillator logic was made in two steps: (1) signals from the 
scintillators in pots 2, 3, and 4 were required in the UP (or DOWN) trigger 
spectrometer, together with the scintillator in pot 1 in veto;
(2) scintillators 2, 3 and 4 in the adjacent DOWN (or UP) spectrometer were 
put in veto. For example for the UP spectrometer, the final scintillator 
logic was the coincidence, 
$\rm (\bar{1} \cdot 2 \cdot 3 \cdot 4)_{UP} \cdot 
(\bar{2} \cdot \bar{3} \cdot \bar{4})_{DOWN}$.
   
This logic suppressed the background events which were mainly due to 
secondary interactions in the quadrupoles originating from beam halo 
particles. This condition is compatible with an inclusive trigger
because the particles emitted in the same arm 
as the fast \alam\ are necessarily low momentum particles 
which could not reach the scintillator Sc2 through quadrupole Q1.

\vspace{2 ex}
\noindent {\bf Multiplicity logic}
\vspace{2 ex}
 
As a further refinement of the trigger, we required that there be two
or three hit wires in chamber 2. Three hits arise quite frequently
because there exists a $10\%$ probability for a particle going
through a wire plane to generate two adjacent hits.
This additional multiplicity logic was necessary because
the Hit Pattern
Logic was based on four-wire groups and therefore could allow events
with up to twelve hit wires, if the hits were distributed among
the twelve wires closest to the beam.
 
This condition was implemented by adding a simple circuit to the TDC
card of the $y$ plane for chamber 2 (for both the spectrometers
above and below the beam in the \ap\ side) which counted the wire
hits as they were being read into the Hit Buffers via the readout
modules.
 
The trigger decisions of Hit Pattern Logic, Scintillator Logic and
Multiplicity Logic were combined in the NIM Trigger Logic to give an
overall \alam\ decision for each of the \ap -spectrometers. The
total time for this trigger decision was about $1.3\,\mu\rm sec$.
 
\vspace{2 ex}
\noindent {\bf Rates and Suppression}
\vspace{2 ex}

The total \alam\ rates for four trigger stages are given 
in Table~\ref{tabsup} for a luminosity of $10^{30}$cm$^{-2}$s$^{-1}$
and for the sum of UP and DOWN spectrometers.
The raw trigger rate is the counting rate
of the scintillator coincidence
$\bar{1} 234$. 
Since we were required by the UA2 data acquisition system to
run with a total UA8 rate of 0.2 Hz, the \alam\ trigger usually used about
one fourth of the total UA8 trigger rate. Roughly 10\% of 
this trigger rate are true \alam\ events.

The data sample collected during two months of the 1989 collider run
corresponds to an integrated luminosity of
$1.5$ pb$^{-1}$. 
During these runs, pots 2 and 3 were positioned 12 beam widths ($\sigma$)
from the beam. Pots 1 and 4 were set at $14\sigma$.
These positions represented a good compromise
between the expected \alam\ event rate
and the additional background rate generated in the UA2
experiment by the presence of the pots.
 
\section{Event Reconstruction}
  
\subsection{Fitting Procedure}
\label{fitproc} 

A $\chi^{2}$ fit is used to select good \alam 's 
from the sample of candidates. 
The \ap\ stays closest to the beam and crosses chambers 2,3,4.
It has its momentum, direction and position
measured. 
The $\pi^+$ crosses chambers 2,3 and has only its direction and position
measured. 
To test the hypothesis that the \ap\ and $\pi^+$ 
come from a particle of mass $M$ which decays inside the quadrupole, Q1,
a ${\chi}^{2}$ minimization is performed using the program MINUIT\cite{minuit}.
There are 7 free parameters: $M$, the \alam\ momentum $P$ and its  
track slopes $dx/dz$, $dy/dz$, the $z_{v}$ coordinate of its decay vertex 
and the decay angles $cos\theta$ and $\phi$.
$\theta$ is the angle of the decay \ap , measured with respect to the
normal to the production plane,
$\hat{n} = \ap _{B} \times \alam $/$\mid \ap _{B} \times \alam \mid$
in the \alam\ rest frame
and $\phi$ is its azimuthal angle with respect to the
\alam\ direction. $\ap _{B}$ and \alam\
are the beam antiproton and \alam\ laboratory
momenta, respectively. 
These seven parameters are determined for each event 
by minimizing the quantity:
\begin{equation}
\chi^{2} \, \, = \, \, 
{{\sum_{i=1}^{30} (c_{i}-c_{0i})^{2}}\over {\sigma_{i}^{2}}}
\label{eq:chisq}
\end{equation}
The $c_{i}$ are the measured hit positions in all chamber planes and the
$c_{0i}$ are the calculated coordinates in the corresponding planes for
the \ap\ and $\pi^+$. 
The $c_{0i}$ are calculated from the above set of seven parameters,
using first order matrix transport equations
to propagate the particles inside the quadrupoles. 
The index i runs from 1 to 30 (5 chamber points times 6 planes per chamber) if
there are no missing hits. $\sigma_{i}$ is the spatial resolution
on a single wire plane position measurement\cite{ua8hard}, 
typically equal to 160 $\mu$.

\begin{figure}
\begin{center}
\mbox{\epsfig{file=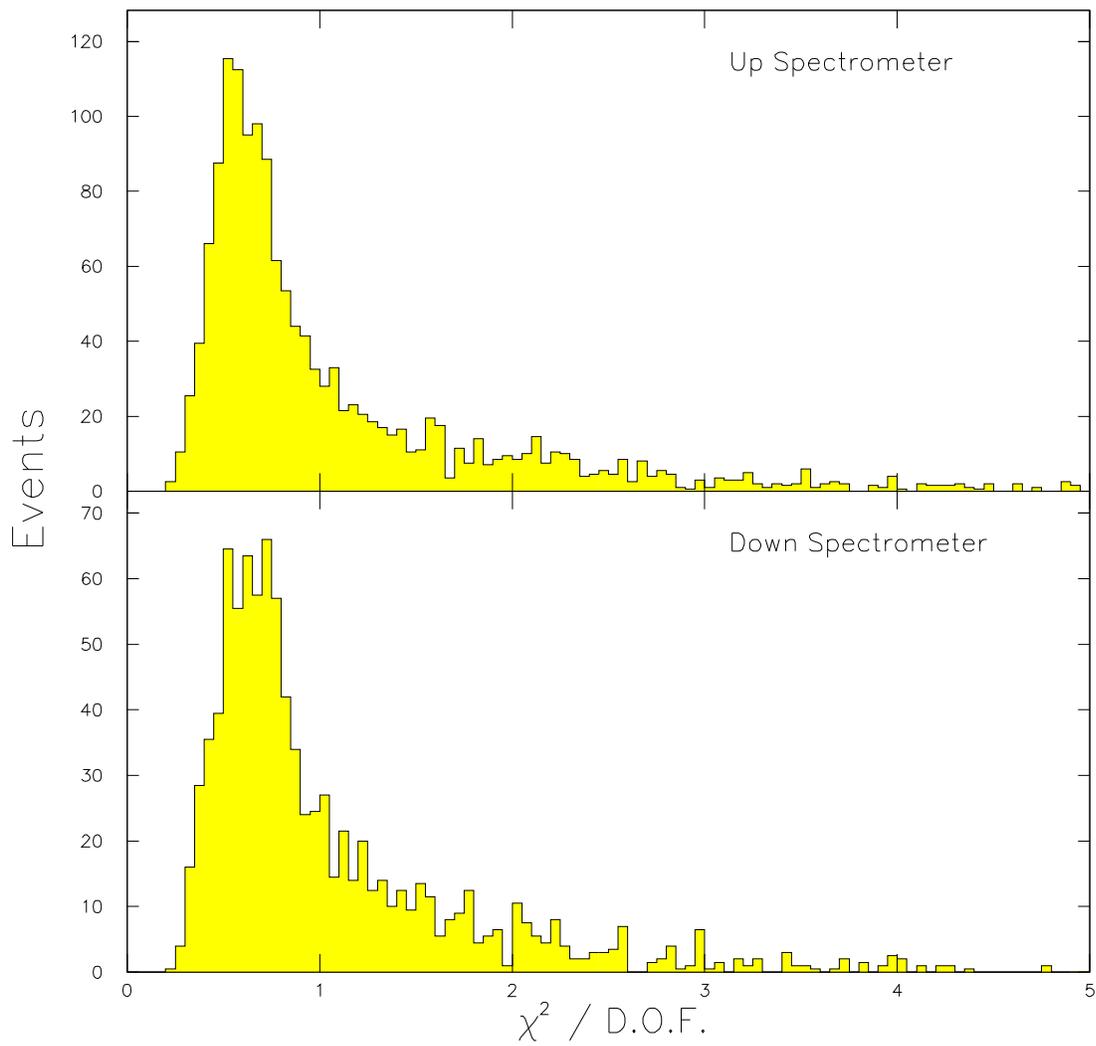,width=16cm}}
\end{center}
\caption[]{
$\chi^{2}$ distributions of the MINUIT fit for events in
the UP and DOWN spectrometers, respectively.
}
\label{fig:lambda_chi2}
\end{figure}

The resulting $\chi^{2}$ distribution, shown in Fig.~\ref{fig:lambda_chi2},
looks reasonable, although there is a tail which extends to quite large values.
The percentage of background events in the distribution is discussed below.
 
There exists one unique solution for about half the events.
The other half have two solutions with equally good ${\chi}^{2}$.
This arises for the following reason; when the \alam\ and its decay
plane are in the vertical plane $y-z$ (see Fig.~\ref{fig:lambda_layout}), the
two decay particles always stay in that plane, because there is no field
component along the $y$ direction inside the quadrupoles. 
The occurence of a 2-solution event depends on the resolution of the 
$x$-coordinate measurement of each point in a chamber and on the systematic 
error on the chamber positioning along $x$. 
Due to the orientation of the stereo views in the chambers, the $x$-resolution 
is much worse than the $y$-resolution: $\sigma _{x}$ $=$ 670 $\mu$ and
$\sigma _{y}$ $=$ 65 $\mu$. The systematic shift along the $x$ direction is 
about 1 mm. 
For both spectrometers, the percentage of 2-solution events is well reproduced 
by Monte Carlo calculations.  

The fits for the 2-solution events are only effective in the $y-z$ plane. In
that case, one solution has the \ap\ going down
and the other solution has the \ap\ going up at the decay vertex point
in the vertical plane $y-z$. 
All variables are stable for the two solutions, except $\phi$ and $z_{v}$. 
The $\phi$ values for the two solutions are related by $\phi _{1}$
$+$ $\phi _{2}$ $\sim$ $0$, and
z by $z_{v1}$ $-$ $z _{v2}$ $\sim$ 70 cm. 
For these events, both solutions are used and each is
weighted by 0.5 in the final \xf , \pt , and 
$cos\theta$ analysis. 
The resolutions on these variables, estimated by
Monte-Carlo calculations, are $\sigma_{\xf} = 0.008$,
$\sigma_{\pt} = 0.009$ GeV and $\sigma_{cos\theta} = 0.1$.
 
\begin{figure}
\begin{center}
\mbox{\epsfig{file=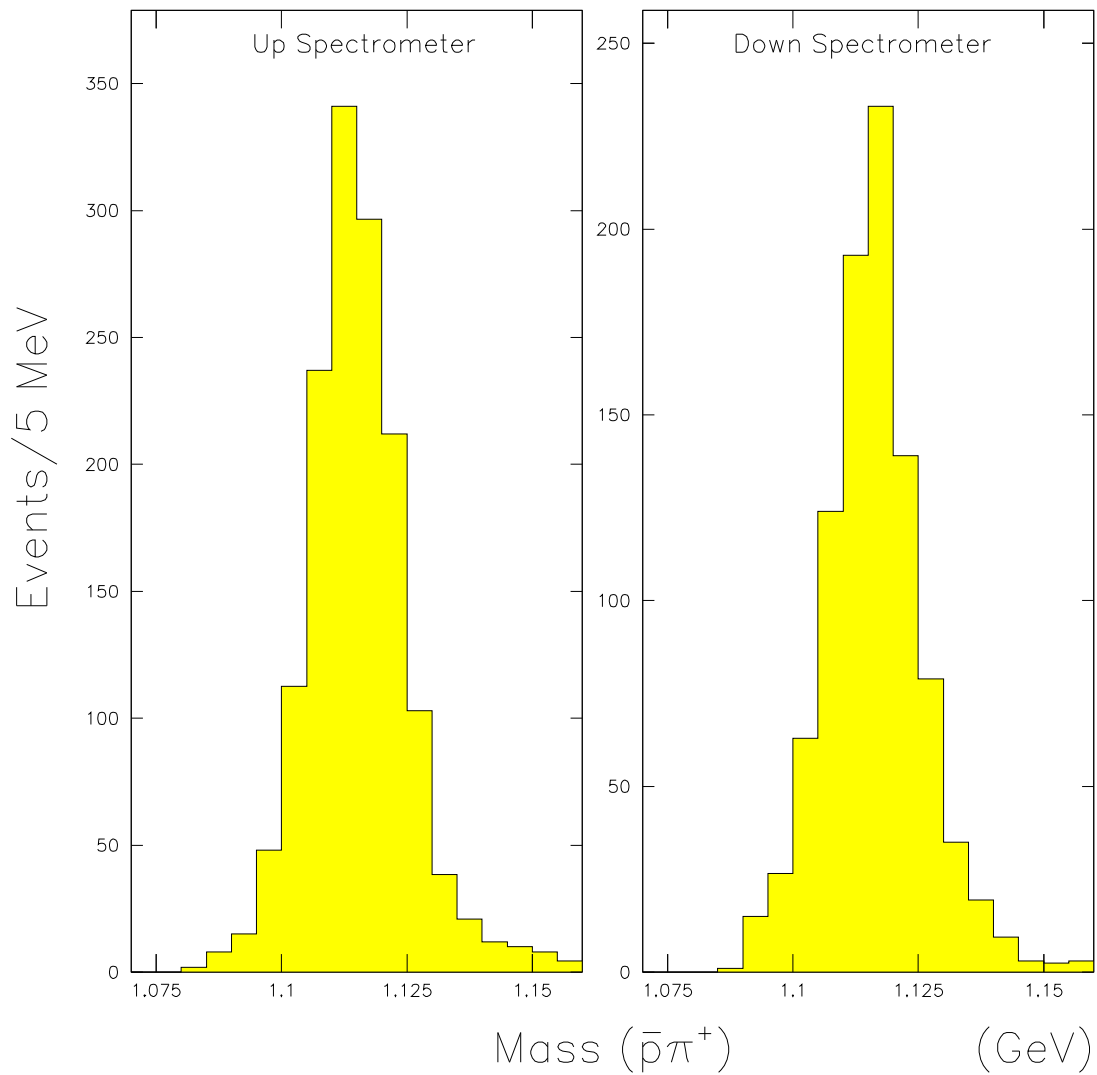,width=16cm}}
\end{center}
\caption[]{
Invariant mass of $\ap \pi^+$ systems measured in 
the UP and DOWN spectrometers, respectively.
}
\label{fig:lambda_mass}
\end{figure}

Figure~\ref{fig:lambda_mass} 
shows the invariant mass distribution of the $\ap \pi^+$
system. It has a well-shaped peak, centered at the
\alam\ mass value.  
A Monte-Carlo calculation shows that the acceptance in the mass range 
between the \alam\ mass and 1.6 GeV is larger
than the acceptance at the \alam\ mass with the same trigger
conditions. Therefore, the mass peak can not be due
to an acceptance effect of our very small aperture spectrometers.

Table~\ref{upsamp} gives a summary of the event samples at each step of the
analysis. ``Chamber points found" means events with
no point in chamber 1, two points each in chamber 2
and 3, and one point in chamber 4.
Reconstructed events are events which are successfully fit with the \alam\
decay hypothesis. There is a final fiducial cut made in the chambers.

\begin{figure}
\begin{center}
\mbox{\epsfig{file=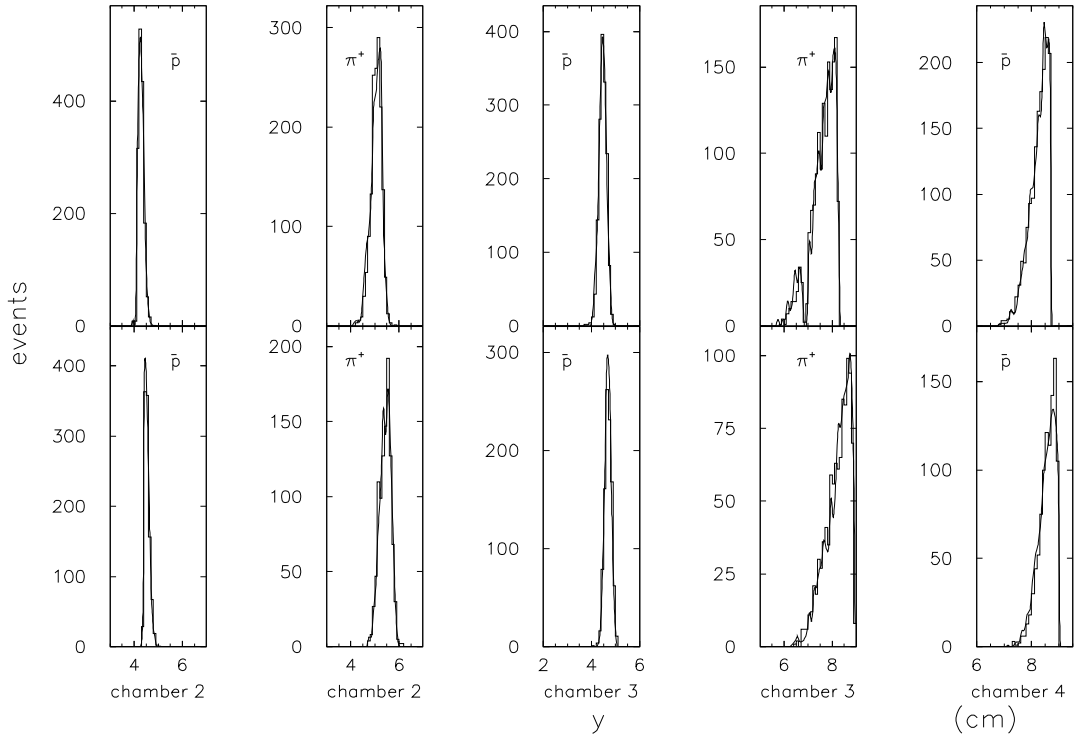,width=16cm}}
\end{center}
\caption[]{
Distributions of the (vertical) $y$-coordinate 
of the \alam\ decay products in all chambers for the UP and DOWN
spectrometers, respectively. Histograms are data and solid curves are 
Monte Carlo calculations. Shifts seen between UP and DOWN spectrometers
are due to different distances from the beam axis for the chambers
in the two spectrometers. 
} 
\label{fig:lambda_y}
\end{figure}

Figure~\ref{fig:lambda_y} shows the $y$-coordinate distributions of the \ap\
and $\pi^+$ in chambers 2,3 and 4 for the UP and DOWN spectrometers,
respectively. The Monte Carlo predictions, discussed is Sect.~\ref{geomacc},
are seen to be in excellent agreement with the data.

\begin{figure}
\begin{center}
\mbox{\epsfig{file=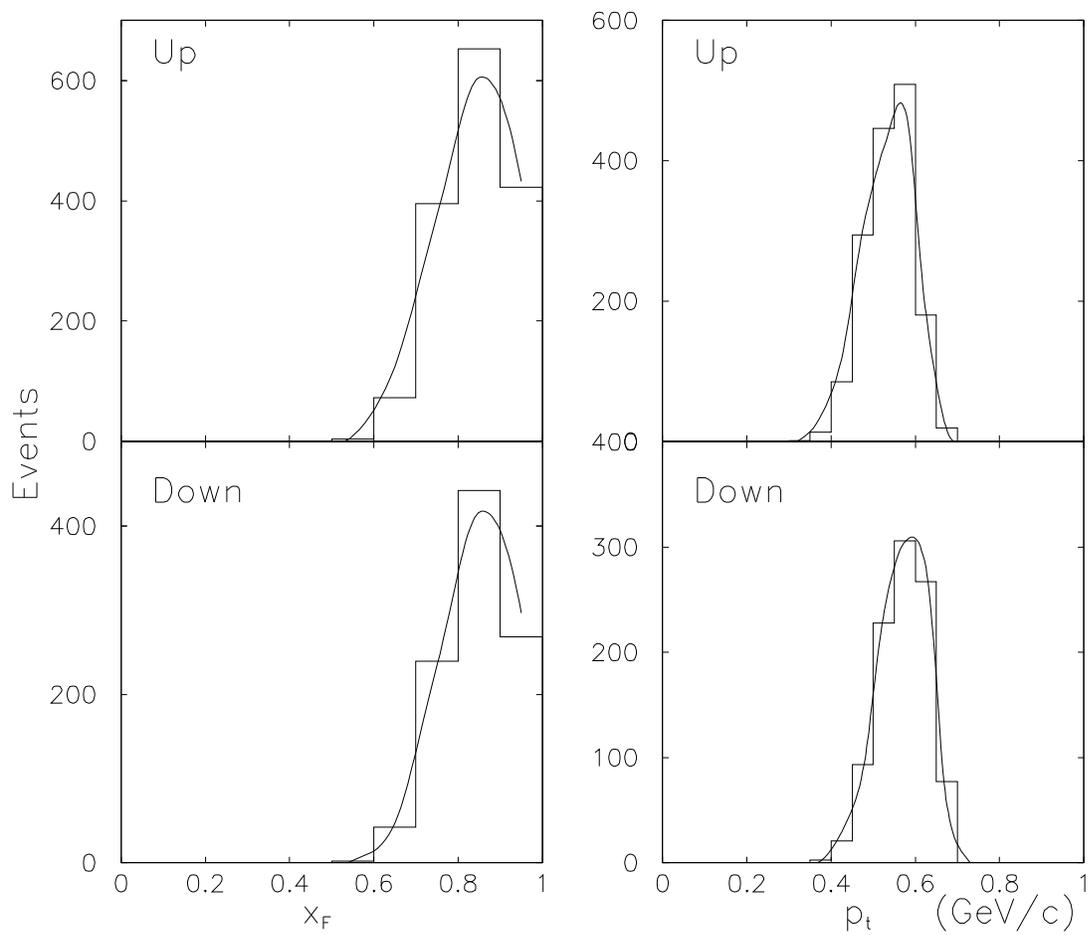,width=16cm}}
\end{center}
\caption[]{
\xf\ and \pt\ distributions of the \alam\ sample
in UP and DOWN spectrometers, respectively. 
The histograms are data and the solid curves are 
Monte Carlo calculations, as explained in the text.
}
\label{fig:lambda_xfpt}
\end{figure}

Figure~\ref{fig:lambda_xfpt} shows the \xf\ and \pt\ distributions of the final
sample of events. They display the ranges covered in these two variables and
illustrate the extreme \xf -values of the data sample.
Again, there is excellent agreement between the Monte Carlo predictions
and the data. 
Since the events reported here are in a rather limited region of
phase space, for all predictions shown in Figs.~\ref{fig:lambda_y}
and \ref{fig:lambda_xfpt}, the Monte Carlo events are generated 
according to the parametrization of the 
measured cross sections of the R608\cite{sherwood} experiment.

\begin{figure}
\begin{center}
\mbox{\epsfig{file=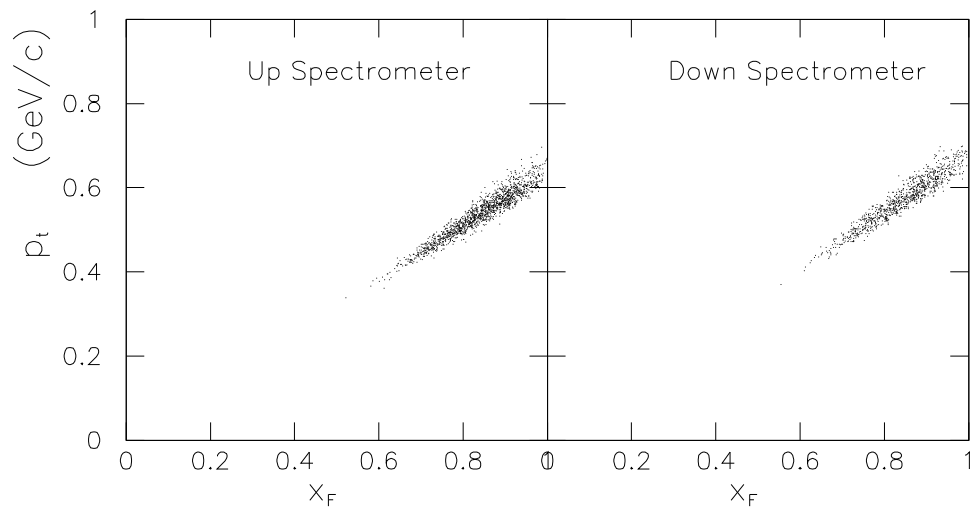,width=16cm}}
\end{center}
\caption[]{
\xf\ vs. \pt\ for the \alam\ sample in 
the UP and DOWN spectrometers, respectively.
}
\label{fig:lambda_scatter}
\end{figure}

Not surprisingly, 
there is a strong correlation between \xf\ and \pt\ due to the small
aperture of the Roman pot spectrometers. This is shown  in
Fig.~\ref{fig:lambda_scatter}, which are scatter plots of the two variables
in the UP and DOWN spectrometers, respectively.

\subsection{Geometric Acceptance}
\label{geomacc}

The geometrical acceptance 
for \alam 's is calculated using a Monte-Carlo simulation of the 
UA8 spectrometer, for \alam\ coming from the crossing
region and decaying near the end of the first quadrupole Q1.
 
The simulation takes into account the parameters which define the
two incident beams: the bunch length,
the angular divergences ($\sim 0.18$ mrad in the
horizontal and vertical planes) and the momentum spread (0.1\%).
The decay particle trajectories inside the quadrupoles are calculated 
using first order matrix transport equations. Multiple scattering
is simulated in the pot windows, chambers and
scintillators. The
digitization of the chamber information takes into account
double hit production, non-linear drift speed and non-uniform TDC
bin sizes.
The efficiency of each chamber was
measured independently and introduced in the calculation. A fiducial
region is defined in each chamber where the efficiency is close
to $100\%$. 
The shape of the pipe boundary inside the quadrupole Q2 limits the 
acceptance of the proton and is carefully simulated 
(see Fig.~\ref{fig:aperture}).
 
Monte Carlo events which satisfy the trigger logic conditions are
processed with the same analysis package used to process the real data.
The multiple scattering calculation has been checked
using an elastic data sample. The momentum
resolution of elastic \ap 's
measured with chambers 2,3,4 is well reproduced by the
Monte-Carlo. 
More generally, the momentum resolution of \ap 's measured by chambers 
2,3,4 is well described between 150 and 315 GeV by the relation
$\sigma_{p} \bar{y}=(\alpha^{2}p^{4}+\beta^{2}p^{2})^{1/2}$,
where $\alpha^{2}=0.74 \times 10^{-7}$ cm$^{2}$GeV$^{-2}$ and
$\beta^{2}=0.20 \times 10^{-2}$ cm$^{2}$. 
$\bar{y}$ is the mean coordinate of the particle trajectory inside
quadrupole Q2 -- the 
momentum resolution depends on the $y$-coordinate, due to the quadrupole
field shape.
For elastic events, $\bar{y}$ and $\sigma_{p}$ values
are equal to $4.8$ cm and $6.6$ GeV, respectively.
 
The measured \pt\ is approximately given by 
$\xf \cdot \ap _{B} \cdot (y/d)$,
where $y$ is the coordinate of the \ap\ in chamber 2 and
$d$ is the distance of this chamber from the crossing region. 
Due to the very small aperture of the spectrometer which accepts
decay protons in a $\Delta y$ interval of only 0.4 cm, the
acceptance is restricted to the very small $\Delta \pt$
range.
For each value of \xf , the acceptance
peaks at a value of \pt\ which depends linearly on \xf .
The acceptance grows with increasing \xf , up to a value about $0.004$.
 
The angular spread of the beams and the beam momentum resolution
smear out measured values of \xf\ and \pt ,
with $\sigma_{\xf}\sim 6\times 10^{-4}$ and $\sigma_{\pt}\sim 0.05$ GeV. 
Therefore, the acceptance in 0.1 bins of \xf\ is nearly independent of the 
beam parameters. 

\subsection{Corrections}

In order to arrive at the absolute cross sections for 
React.~\ref{eq:alaminclusive}, we calculate additional
corrections, whose effects are not included 
in the Monte-Carlo simulations described above. 
All these corrections with their 
estimated errors are listed in Table~\ref{icorr}.
 
The hit patterns in the chambers have been carefully
simulated in the Monte-Carlo calculations, taking into account 
double hit production, the multiplicity trigger logic
requirement and efficiencies. 
However, the Monte Carlo does not 
take into account any extra hits in the chambers. 
Since \alam\ candidates which contain at least
one extra point in a chamber are rejected during the 
analysis, this effect was corrected for statistically.
The fraction of events lost was estimated using a sample of
events taken without these requirements. 

A sample of such rejected data events which contain three points
in chamber 2, two points in chamber 3 and one point in chamber 4 were
studied in order to get an upper estimate ($3.4\%$)
of \alam\ events lost. Other rejected patterns with
an extra point in chambers 3 or 4 are much less frequent
and the corresponding corrections can be neglected.
 
The relative positions of chambers 3 and 4 with respect to chamber 2 
are determined to within $\pm 50 \, \mu$,
using the momentum distribution of the elastic
antiproton and also the \alam\ mass distribution.
Assuming a $\pm1$ mm shift in the nominal value of the $y$ position of 
chamber 2 in the Monte Carlo, we found a systematic variation in the 
number of accepted events by $\pm 15\%$. The systematic error introduced on 
the acceptance calculation due to $x$ misalignment of the chambers relative
to nominal beam axis is estimated to be $\pm 3.5\%$

Monte-Carlo calculations were done with a nominal vacuum pipe. 
If the dimension of the pipe is reduced by $1$ mm, the
acceptance value changes by $\sim 15 \%$.
That change corresponds to the maximum possible deviation from the
nominal value tolerated during the design and installation
inside the quadrupoles.
 
As noted above, the $\chi^{2}$ distribution (see Fig.~\ref{fig:lambda_chi2})
has a small tail at large $\chi^{2}$ values. 
The $\chi^{2}$ distribution can be understood 
if systematic uncertainties in the transverse ($x$) chamber positions are
taken into account.
A correction
factor of 0.94 is included in Table~\ref{icorr} to take into account the
presence of background events in the final sample.

The loss of events due to particle interactions in the
windows of the pots, scintillators and chambers have also been estimated.
Other corrections are for the decay probability of the $\pi^+$ 
and the unseen decay mode of the \alam . 
We also multiply the resulting cross section by a factor of 2 to 
compensate for the unseen \lam\ particles produced in the opposite hemisphere,
thus permitting a direct comparison with 
previous measurements of \lam -production in pp interactions.
The uncertainty in the luminosity measurement is obtained
by comparing two different monitors.

Multiplication of all these correction factors gives us 
an overall correction of 4.04 which is applied to the data. 
The systematical uncertainty on the absolute cross sections 
is estimated to be $\sim \pm 20\%$. 

\section{Physics Analysis}

\subsection{Cross Section}
 
\begin{figure}
\begin{center}
\mbox{\epsfig{file=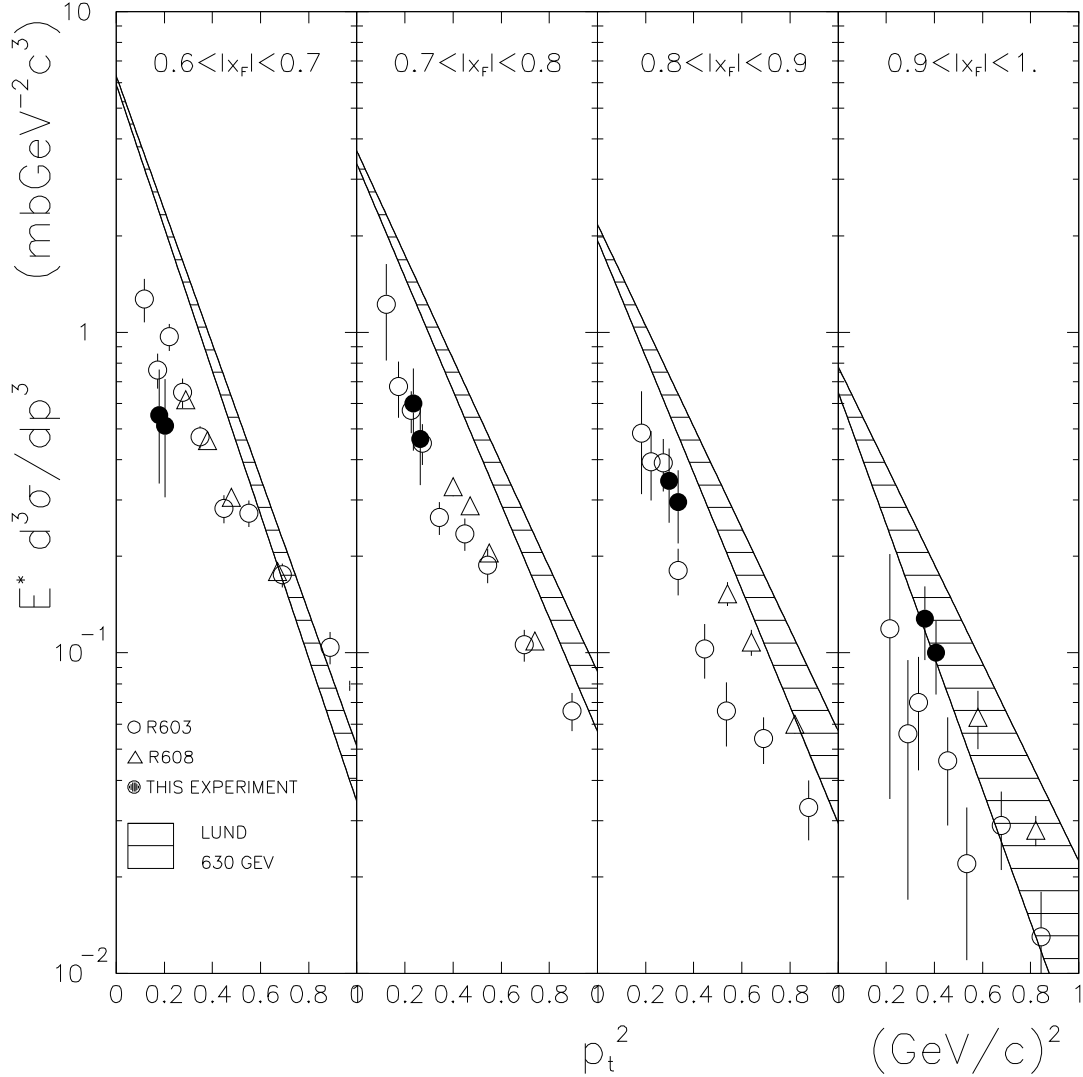,width=16cm}}
\end{center}
\caption[]{
Measured invariant \alam\ cross section in four bins of \xf\ 
versus $p_t^2$ compared with previous measurements from the R603
\protect\cite{erhan} and R608\protect\cite{sherwood} experiments at the ISR.
The cross sections are multiplied by a factor of 2, so that they can be 
directly compared with the 2-arm total \lam\ cross section in pp interactions.
The bands for each bin of \xf\ show the PYTHIA Monte-Carlo calculated values
discussed in the text. The hatched regions represent
$\pm 1 \sigma$ regions for the Lund predictions at 630 GeV.
}
\label{fig:lambda_xsecm}
\end{figure}

Table~\ref{icro} shows the invariant
cross sections in four bins of \xf , for the UP and DOWN 
spectrometers separately. In each case,
the first error is statistical and the 
second error is the systematic uncertainty.
Together with the results of this experiment, Fig.~\ref{fig:lambda_xsecm} shows
the previous measurements made at the CERN-ISR by the R603\cite{erhan} and
R608\cite{sherwood} experiments. 
Although, there are systematic differences seen between the 
R603 and R608 data sets for $\xf > 0.8$, 
our present results are compatable with the 
(second generation) R608 points
over the entire \xf -range shown.
No significant increase of the invariant cross section is seen in the
center-of-mass energy range, 52--630~GeV.

\subsection{Lund Monte-Carlo calculations}
 
The invariant cross sections for React.~\ref{eq:alaminclusive}
are compared in Fig.~\ref{fig:lambda_xsecm}
with Lund Monte-Carlo calculations 
for $p \ap $ interactions at 630 GeV. The default
values of the programs (PYTHIA 5.6 and JETSET 7.3)\cite{lund} were
used throughout these calculations. 
The comparison shows that  
the Lund Monte Carlo values of the
invariant cross section are 
2-3 times larger than the measured ones. 
This discrepancy is most likely due to our limited knowledge 
of the baryon fragmentation function\cite{sjos} used in PYTHIA.

\subsection{\alam\ Polarization}

\begin{figure}
\begin{center}
\mbox{\epsfig{file=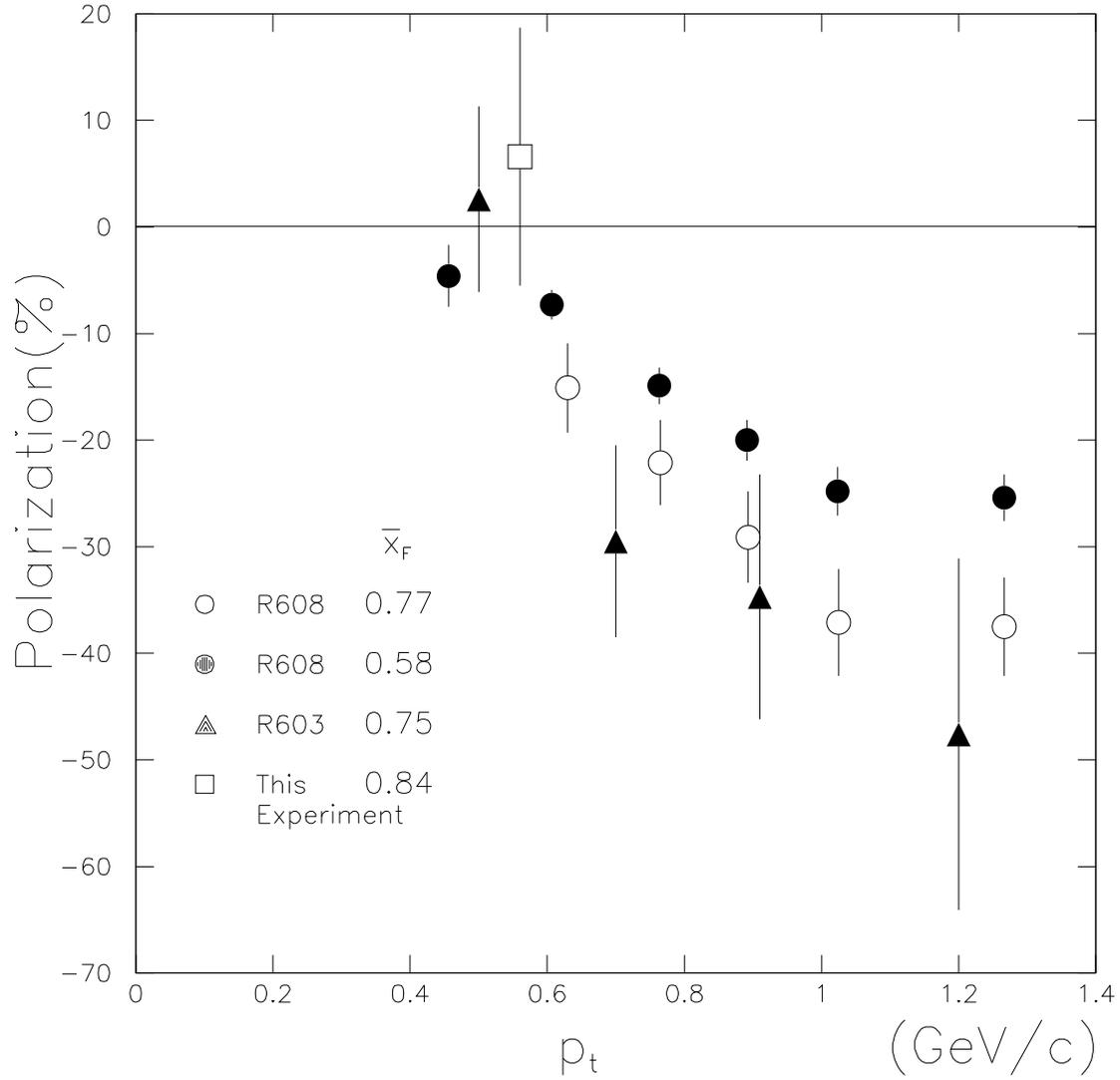,width=16cm}}
\end{center}
\caption[]{
Measured $\Lambda$ polarization together with previous
measurements from R603 \protect\cite{erhanpol} and R608\protect\cite{sherwood}
shown versus $p_t$. The result from our experiment is averaged over
UP and DOWN spectrometers. The mean $\bar{\xf}$ values are shown 
for each set of data.
}
\label{fig:lambda_pol}
\end{figure}

The \alam\ polarization, $\cal{P}$, has been obtained from the $cos\theta^{'}$
distribution, where $\theta^{'}$ is the angle of the emitted \ap\ 
in the \alam\ rest frame, 
with respect to the final (precessed) polarization vector.
Because the precession angle 
is small (a mean value of $6^{\circ}$) 
$\theta^{'}$ does not differ very much from $\theta$, the measured $\theta$
defined in Sect.~\ref{fitproc}.
The raw distribution
is modified to take into account the acceptance correction factor
and is fitted according to the formula
$dN/dcos\theta^{'}=K(1+\alpha {\cal P} cos\theta^{'})$ where
$\alpha = -0.642$ is the \alam\ decay parameter and $cos\theta^{'}$ is
in the range, -0.8 to 0.8. Outside this range, 
$cos\theta^{'}$
is not reliably calculated due to the unavoidable uncertainty
in the chamber positioning along $x$. 
The measured $\cal{P}$ values are
$0.00 \pm 0.15$ and $0.17 \pm 0.20$, in the UP and DOWN
spectrometers, respectively. 
These results are consistent with previous
measurements done at the ISR\cite{erhanpol,sherwood}
in the same \xf\ and \pt\ ranges, as shown in Fig.~\ref{fig:lambda_pol}.

\section{Conclusions}
 
We have reported a measurement of the inclusive \alam\ 
invariant cross section in the \xf\ range, 0.6-1.0. 
The observed \alam 's are emitted forward at a very small angle
with respect to the outgoing \ap\ beam ($\sim 2$ mrad)
and decay inside a quadrupole in a 2m length region which
starts 19m from the interaction region.
They are clearly identified using a $\chi^2$ minimization procedure.
 
The cross sections are found to be indistinguishable from previous measurements
in the fragmentation region made at the ISR in the 
c.m. energy range, 52-62 GeV. 
The polarization of the \alam 's is found to be consistent with 
zero, as it was for \lam 's in the same \xf\ and \pt\ region,
produced in $pp$ interactions 
at the lower ISR energies.

Finally, we have shown that PYTHIA Monte Carlo predictions for \alam\
production at large \xf\ are a factor of 2 to 3 larger than our measurements.
Our results can be used to ``tune" PYTHIA to give 
more reliable baryon predictions in the fragmentation region.

\section*{Acknowledgements}

We are grateful to the CERN administration for their continued support.
We also thank the UA2 collaboration for allowing us to use their
data acquisition system for the measurements reported here.
M.T.Z. thanks the Scientific \& Technical Research Council of Turkey
(TUBITAK) for support as a Ph.D. Fellow during his stay at CERN. 
A.K. and N.O. wish to thank Cukurova University, TUBITAK and ICSC-World
Lab for their support.

\pagebreak

\clearpage

\begin{table}
\centering
\begin{tabular}{|c|c|}                            
\hline
Trigger Requirement           & Rate (Hz)         \\ 
\hline
Scintillators (Raw)             & 400.            \\
Hit pattern Logic               & 1.              \\
Scintillator Logic              & 0.1             \\
Multiplicity Logic              & 0.05            \\ 
\hline
\end{tabular}
\caption{\alam\ trigger rate suppression. See text for a description of
the three trigger components.}
\label{tabsup}
\end{table}

\begin{table}[htbp]\centering
\vspace{0.7cm}
\begin{tabular}{|l||c|c|}                                        \hline
                      &UP             &DOWN \\ 
Event Category    &Spectrometer  &Spectrometer \\ \hline
Raw Triggers         &40072  &35890 \\
Chamber points found &10319  &7481  \\
Reconstructed \alam\ &5614   &2348  \\
After fiducial Cut   &1548   &995   \\ \hline
\end{tabular}
\caption{\alam\ event sample. See text for a discussion of the reconstruction
steps. The unequal numbers of events in UP and DOWN spectrometers is due to
different distances of the chambers in the two spectrometers from the beam
axis.}
\label{upsamp}
\end{table}

\begin{table}[htbp]\centering
\vspace{0.7cm}
\begin{tabular}{|l|c|}                                             \hline
Correction               &Factor                     \\   
\hline
Background in $\chi^{2}$ dist. &0.940$\pm$0.010         \\
Chamber 1 Veto            &1.190$\pm$0.054         \\
Extra point in chamber 2  &1.034$\pm$0.005         \\
Pipe                      &1.075$\pm$0.075         \\
Nuclear Interactions      &1.041$\pm$0.007         \\
Decay of $\pi^+$        &1.002                   \\
Unseen decay mode $\alam \ra n\pi^{0}$  &1.558     \\
Unseen Hemisphere         &2.000                   \\
$x$-coordinate              &1.000$\pm$0.035         \\
$y$-coordinate              &1.000$\pm$0.150         \\
Luminosity measurement error &1.000$\pm$0.081      \\ 
\hline
FINAL CORRECTION          &4.042$\pm$0.812         \\ 
\hline
\end{tabular}
\caption{Cross Section Corrections}
\label{icorr}
\end{table}

\begin{table}[htbp]\centering
\begin{tabular}{|c|c|c|c||c|}                                      
\hline
\xf\   &$<\xf>$  &$<\pt^2>$   &Events &E$^{\ast} d^3 \sigma /dp^3$ \\
bin               &         &(GeV$^2$) &       &(mb GeV$^{-2}$)    \\ 
\hline
\hline
\multicolumn{5}{|c|}{UP Spectrometer} \\ 
\hline                          
0.6-0.7    &0.67          &0.18 &73   &0.552$\pm$0.107$\pm$0.107\\
0.7-0.8    &0.76          &0.19 &396  &0.599$\pm$0.055$\pm$0.116\\
0.8-0.9    &0.85          &0.30 &653  &0.344$\pm$0.023$\pm$0.066\\
0.9-1.0    &0.935         &0.36 &423  &0.128$\pm$0.079$\pm$0.025\\ 
\hline 
\hline 
\multicolumn{5}{|c|}{DOWN Spectrometer} \\ 
\hline 
0.6-0.7    &0.67          &0.20 &43   &0.510$\pm$0.105$\pm$0.099\\
0.7-0.8    &0.76          &0.26 &240  &0.464$\pm$0.041$\pm$0.090\\
0.8-0.9    &0.85          &0.33 &443  &0.295$\pm$0.019$\pm$0.057\\
0.9-1.0    &0.94          &0.40 &269  &0.100$\pm$0.068$\pm$0.019\\ 
\hline
\end{tabular}
\caption[]{
Invariant cross sections for \alam\ in 
React.~\protect\ref{eq:alaminclusive}, multiplied by a factor of 2 to allow
comparison with 2-arm inclusive \lam\ cross section in 
React.~\protect\ref{eq:laminclusive}.
}
\label{icro}
\end{table}

\clearpage
  
\clearpage

\clearpage

\clearpage

\clearpage

\clearpage

\clearpage

\clearpage

\clearpage

\clearpage

\end{document}